# Design of an Adaptive Surface for Space-Reconfigurable Reflectors using Auxetic Lattice Skin

Bin Xu[1], Houfei Fang[2], Yangqin He[2], Shuidong Jiang[2], Lan Lan[2,]
1 China Academy of Space Technology(Xi'an), Xi'an 710000, China
2 Shanghai YS Information Technology Co., Ltd., Shanghai, P. R. China

## Abstract

The effect of Poisson's ratio to the reflector reshaping is investigated through mechanical study of reconfigurable reflectors in this paper. The value of Poisson's ratio corresponding to the minimum deforming stress is given and an auxetic lattice is proposed for the reflector surface. The parameters of the auxetic lattice are investigated for vary Poisson's ratio. A case of reconfigurable reflector is studied, the curvature change and strain are calculated by surface geometry analyse, and the negative Poisson's ratio is established for vary thickness. According to RMS calculation by the FEM structure analyse, the thickness can finally be established.

## 1 Introduction

Conventional reflecting antennas with fixed reflector surface can not adapt the radiation patterns. In contrast, the in-orbit space antenna with mechanically reconfigurable reflector(MRR) can adapt its radiation pattern by reshaping the reflector,and cover several different areas during lifetime. So a space-reconfigurable antenna can reduce the quantity of antennas and saving launching cost. In addition, MRR can reduce the manufacturing accuracy by compensating their shape errors. Thus, a MRR has potential for a variety use of space applications such as Earth Observations and Telecommunications.

In past few decades,many investigations have been performed in MRR research[1-18]. As a pioneer, Prof.P.J.B.Clarricoats[1-4] investigated the reshaping of a metal tricot mesh with a number of actuators distributed over the surface. According his work, for achieving smooth reshaping, the ratio of bending stiffness to tensile modulus (D/E) should be large enough.

K.Photopidan skip the tricot mesh and use a net of interwoven wires to increase the bending stiffness.In the work of Ref. 5And Ref.7, the ends of the wires can free glide through the hole at the fixed rim.Thus,the tensile stress became small, and the effective stiffness ratio D/E was high. In Ref.6, the rim of the reflecting surface was released, and no requirement for tensile modulus, see Fig.1(a).

Leri Datashvili etc.[16,17] designed a reflector morphing skin using flexible fiber composites, which have good mechanical and radio frequency performances by using particle filled silicone as a front layer of the laminate, and a prototype driven by 19 actuators was fabricated, see Fig.1(b). Ref.16 assessed three candidate materials to be used as reflecting surface: biaxially woven fabric carbon fibre reinforced silicone(BWF CFRS), triaxially woven fabric carbon fibre reinforced silicone(TWF CFRS) and orthogonal woven wire meshes with different diameter wires. TWF CFRS was excluded for high actuator forces and manufacturing complexities. And combined orthogonal wire mesh has lowest actuator forces with similar accuracy performance as BWF CFRS.

Shubao Shao with coworkers[18] employed a sandwich structure material to fabricate a reflector surface for its construction simplicity and flexibility in adjusting the ratio of bending stiffness to tensile modulus(D/E). That sandwich structure is composed of three layers: BWF CFRS layer with orthogonally woven fibers on the top and bottom, aluminum honeycomb as the core layer, as shown in figure .

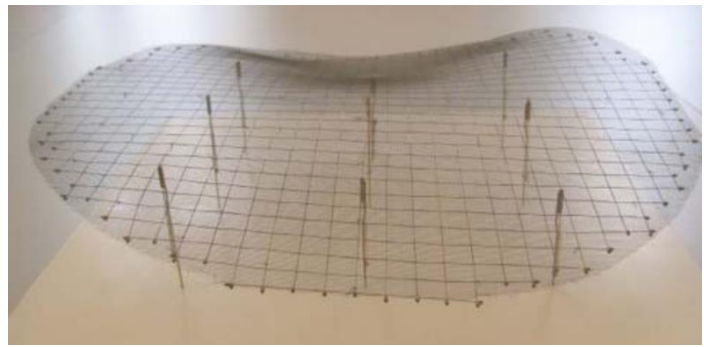

(a)

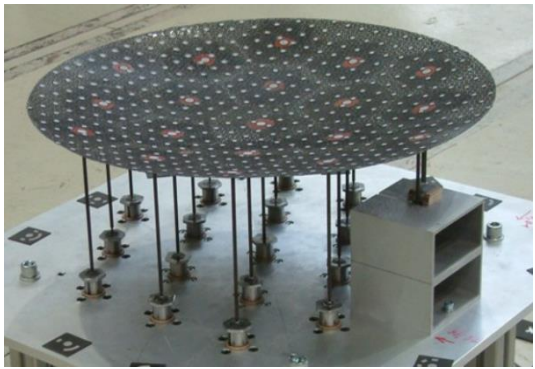

(b)

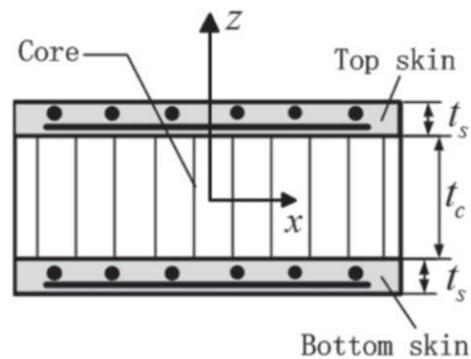

(c)

Fig.1 The morphing skin of MRR

The previous researches of mechanical properties needed for the reconfigurable surface are most about tensile modulus, shear modulus and bending stiffness. As Poisson's ratio is unchangeable for a certain material, most of which are positive, researchers have little interest in the investigation of the effect of Poisson's ratio on surface reshaping. During the last few years, so-called "designer materials" with arbitrarily complex nano/micro-architecture have attracted increasing attention to the concept of mechanical metamaterials[19]. Owing to their rationally designed nano/micro-architecture, mechanical metamaterials exhibit unusual properties at the macro-scale, such as various Poisson's ratio, including negative Poisson's ratio(NPR).

In the present paper, the stress analyses on reflector reshaping is performed, and the relationship between the Poisson's ratio and main principal stress on reflector surface during reshaping is investigated. In the situation of large curvature deformation, material with negative Poisson's ratio has advantage for its small stress, and a NPR lattice skin is introduced to form the reflector surface.

# 2 Mechanical investigation of Surface reshaping of reflectors surface

The reflector reshaping is achieved by the actuation of numbers of actuator under the surface. A elastic out-of-plane deformation from original shape is achieved through a combination of a change in curvature and in-plane strain. And the stress is found by the superposition of bending($\sigma_B$) and membrane (in plane) stress($\sigma_M$). According to the theory of elastic mechanics, the following formulas are used to calculate the stress of the morphing skin in the x ,y direction and shear stress.

$$\sigma_x = \frac{E}{1-v^2}(\varepsilon_x + v\varepsilon_y) \quad (1a)$$

$$\sigma_y = \frac{E}{1-v^2}(\varepsilon_y + v\varepsilon_x) \quad (1b)$$

$$\tau_{xy} = \frac{E}{2(1+v)}\gamma_{xy} \quad (1c)$$

in which $E$ is the Young's modulus, $v$ is the Poisson's ratio of the material, $\gamma_{xy}$ is the shear strain, and $\varepsilon_x, \varepsilon_y$ are strain in dimensions $x$ and $y$.

## 2.1 Stress analysis of bending deformation dominated

For the case of a reflector surface, $K_x$ and $K_y$ are vary with position or with actuation. And $K_x \times K_y$ may be either positive which appears a bowl shaped surface, or negative, which corresponds to a saddle shape. In the situation of bending deformation dominated, the membrane stress is neglected, and consider the principal curvature directions as $x$ and $y$ dimension. Thus, the major principal stress is the maximum of $\sigma_{Bx}$ and $\sigma_{By}$. The bending strain on surface in two direction can be calculated by curvature change ($K_x, K_y$) and thickness $t$:

$$\varepsilon_{Bx} = \delta K_x \times \frac{t}{2} \quad (2a)$$

$$\varepsilon_{By} = \delta K_y \times \frac{t}{2} \quad (2b)$$

So the bending stress in $x$ and $y$ dimension are :

$$\sigma_{Bx} = \frac{Et}{2(1-v^2)}(\delta K_x + v\delta K_y) \quad (3a)$$

$$\sigma_{By} = \frac{Et}{2(1-v^2)}(\delta K_y + v\delta K_x) \quad (3b)$$

In order to investigate the the effect of Poisson's ratio on stress, Assume:

$$\beta = \frac{\delta Ky}{\delta Kx}, \quad \delta Kx \neq 0 \tag{4}$$

And equation(3) can be write as:

$$\sigma_{Bx} = \frac{Et\delta K_x}{2} \frac{1+\beta v}{(1-v^2)} \tag{5a}$$

$$\sigma_{By} = \frac{\alpha \delta K_x}{2} \frac{\beta + v}{(1-v^2)} \tag{5b}$$

The maximum principal bending stress:

$$\sigma_B = \max\left\{\frac{Et\delta K_x}{2}\frac{1+\beta v}{1-v^2}, \frac{Et\delta K_x}{2}\frac{\beta+v}{1-v^2}\right\} \tag{6}$$

$Et\delta K_x/2$ can be considered as a proportionality coefficient $K$, and the investigation on the relationship between the bending stress with Poisson's ratio come down to $\frac{1+\beta v}{(1-v^2)}$ and $\frac{\beta+v}{(1-v^2)}$. As the domain of Poisson's ratio is $(-1,1)$, $\sigma_B$ can also be expressed as a piecewise function:

$$\sigma_B = \begin{cases} K\dfrac{-\beta-v}{1-v^2}, & \beta \in (-\infty,-1) \\ K\dfrac{1+\beta v}{1-v^2}, & \beta \in [-1,1] \\ K\dfrac{\beta+v}{1-v^2}, & \beta \in (1,+\infty) \end{cases} \tag{7}$$

In order to find the maximum bending stress at vary Poisson's ratio, derivative of $\sigma_B$ is calculated as followed:

$$\frac{d\sigma_B}{dv} = \begin{cases} K\dfrac{-v^2-2\beta v-1}{(1-v^2)^2}, & \beta \in (-\infty,-1) \\ K\dfrac{\beta v^2+2v+\beta}{(1-v^2)^2}, & \beta \in [-1,1] \\ K\dfrac{v^2+2\beta v+1}{(1-v^2)^2}, & \beta \in (1,+\infty) \end{cases} \tag{8}$$

Assuming $\dfrac{d\sigma_B}{dv} = 0$, the value of $v$ corresponding to the minimum bending stress (short for "$v_{B\min}$") is followed. And when $\beta = \pm 1$, the bending stress is monotonic about Poisson's ratio. Taking the domain of $v$ into consideration, the value over the range of $(-1,1)$ should be excluded.

$$v_{B\min} = \begin{cases} -\beta - \sqrt{\beta^2-1} > 0, & \beta \in (-\infty, -1) \\ \dfrac{-1+\sqrt{1-\beta^2}}{\beta} > 0, & \beta \in (-1,0) \\ 0, & \beta = 0 \\ \dfrac{-1+\sqrt{1-\beta^2}}{\beta} < 0, & \beta \in (0,1) \\ -\beta + \sqrt{\beta^2-1} < 0, & \beta \in (1,+\infty) \end{cases} \quad (9)$$

According to the above analysis, when $\beta < 0$, $v_{B\min}$ is positive. On the opposite, $v_{B\min}$ is negative. And $v_{B\min}$ is zero when $\beta = 0$. To verify the above conclusions, the discrete values of $\beta$ investigated are given below:

$$\beta = \{-5, -3, -1, 0, 1, 3, 5\}$$

The curves of bending stress vs. Poisson's ratio at different $\beta$ are shown in Fig.2. When $\beta = -5, -3, 0, 3, 5$, $v_{B\min}$ are 0.101, 0.172, 0, -0.172, -0.101 respectively. When $\beta = 1$, the bending stress is increase with $v$, and when $\beta = -1$, the bending stress is decrease with $v$.

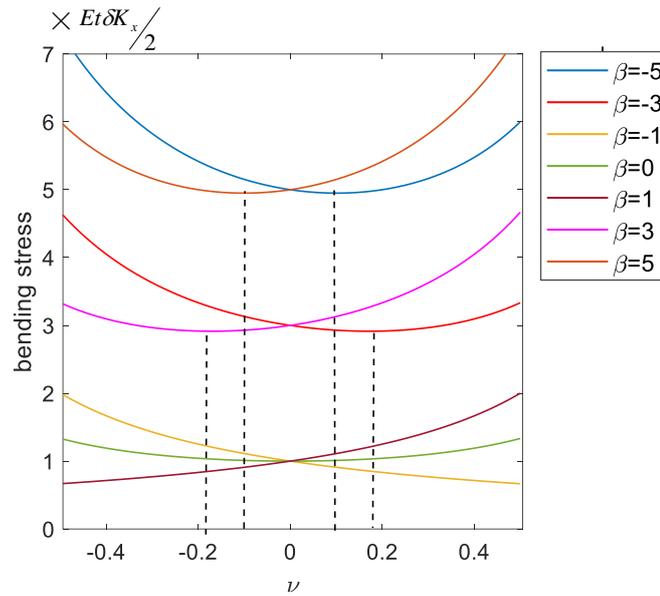

Fig.2 The curves of bending stress vs. Poisson's ratio

## 2.2 Membrane stress vs. Poisson's ratio

The membrane stress in *x* and *y* dimension are as followed:

$$\sigma_{Mx} = \frac{E}{1-v^2}(\varepsilon_{Mx} + v\varepsilon_{My}) \quad (10a)$$

$$\sigma_{My} = \frac{E}{1-v^2}(\varepsilon_{My} + v\varepsilon_{Mx}) \quad (10b)$$

The main principal stress for the situation of only in-plane deformation is

$$\sigma_M = \max(\sigma_{Mx}, \sigma_{My}) \tag{11}$$

Similarly as the investigation of bending stress, when $\varepsilon_{xm}/\varepsilon_{ym} < 0, \neq -1$, the Poisson's ratio corresponding to the minimum membrane stress(short for "$v_{Mmin}$") is positive, when $\varepsilon_{xm}/\varepsilon_{ym} > 0, \neq 1$, $v_{Mmin}$ is negative. When $\varepsilon_{xm}/\varepsilon_{ym} = \pm 1$, the membrane stress is monotonic increase or decrease with Poisson's ratio. And if one of $\varepsilon_{xm}$, $\varepsilon_{ym}$ is zero, the Poisson's ratio corresponding to the minimum membrane stress is zero.

## 2.3 Stress analysis of reflector reshaping

The deformation of reflector surface under point actuation mostly including bending and extension. Total strain in $x$ and $y$ dimension combined bending strain and membrane strain can be add together in these form:

$$\varepsilon_x = \varepsilon_{Bx} + \varepsilon_{Mx} \tag{11a}$$

$$\varepsilon_y = \varepsilon_{By} + \varepsilon_{My} \tag{11b}$$

Combining formula (1),(2),(11), total stress in $x$ and $y$ dimension and shear stress can be calculated by following matrix:

$$\begin{bmatrix} \sigma_x \\ \sigma_y \\ \tau_{xy} \end{bmatrix} = \begin{bmatrix} \dfrac{E}{1-v^2} & \dfrac{vE}{1-v^2} & 0 \\ \dfrac{vE}{1-v^2} & \dfrac{E}{1-v^2} & 0 \\ 0 & 0 & \dfrac{E}{2(1+v)} \end{bmatrix} \begin{bmatrix} \varepsilon_{Mx} + \dfrac{t}{2}\delta k_x \\ \varepsilon_{My} + \dfrac{t}{2}\delta k_y \\ \gamma_{xy} \end{bmatrix} \tag{12}$$

The main principal stress is obtained by bring the value of $\sigma_x$, $\sigma_y$ and $\tau_{xy}$ into equation (13):

$$\sigma_1 = \frac{1}{2}|\sigma_x + \sigma_y| + \frac{1}{2}\sqrt{(\sigma_x - \sigma_y)^2 + 4\tau_{xy}^2} \tag{13}$$

## 3 Geometry analysis of reflector surfaces

The reflector prototype presented in this paper has two shapes which are corresponding to two different coverage area(area 1,area 2), as shown in Fig.3(a),and the displacement is shown in Fig.3(b). Matlab is used to calculate the curvature in $x$ and $y$ dimension and

Gaussian curvature by equation(14). The curvature change of surface reshaping from shape 1 to shape 2 are shown in Fig.4.

$$K_x(x, y) = \frac{\partial^2 z / \partial x^2}{(1+(\frac{\partial z}{\partial x})^2)^{3/2}} \tag{14a}$$

$$K_y(x, y) = \frac{\partial^2 z / \partial y^2}{(1+(\frac{\partial z}{\partial y})^2)^{3/2}} \tag{15b}$$

$$Kg = K_x \times K_y \tag{15c}$$

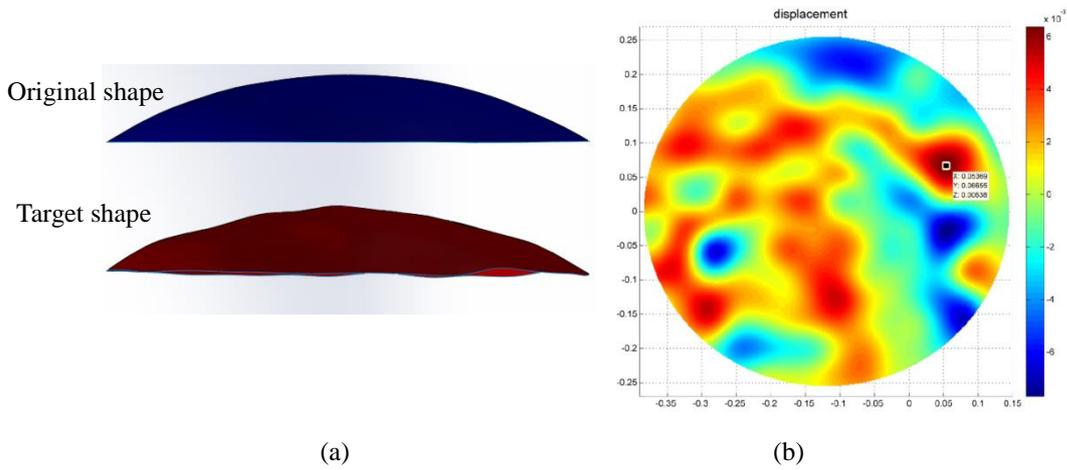

(a)          (b)

Fig. 3 Surface analysis (a) original shape and target shape (b) the distribution of Gaussian curvature

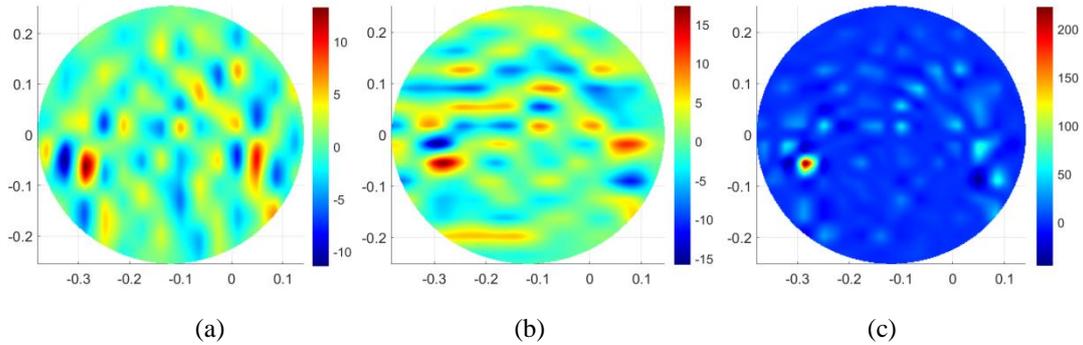

(a)          (b)          (c)

Fig.4 Curvature calculation (a) change of $K_x$ (b) change of $K_y$ (c) change of $K_g$

In order to reduce the stress at rim, the boundary defined for the proposed reconfigurable reflector is partially fixed rim with free radial displacements only. The maximum stress occurs at point (short for "point P") of the maximum Gaussian curvature change because the strain at this point is maximum. Reducing the maximum stress can increase the allowance provided by material yield stress.

As can be seen from the calculation results of Fig. ,the curvature change at point P in x and y dimension is $\delta K_x = 12.67, \delta K_y = 17.36$.

The membrane strain at point P can be calculated by the surface geometry deformation,as shown in Fig..Where $P_1, P_2$ are the nodes at shape 1 and shape 2 of reflector surface corresponding to point P respectively. Point $A_1, A_2$ are the nodes next to point P at shape 1 and shape 2 in $y$ dimension. Point $B_1, B_2$ are the nodes next to point P at shape 1 and shape 2 in $x$ dimension. The membrane strain in $x$ and $y$ dimension can be approximately calculated as shown in Fig.5. Where $L_{P_1A_1}, L_{P_2A_2}, L_{P_1B_1}, L_{P_2B_2}$ are the length of $P_1A_1, P_2A_2, P_1B_1, P_2B_2$ respectively. $\gamma_{xy}$ is the shear strain, and $\alpha, \beta$ are the angles between line $P_1A_1$ and $P_2A_2$, line $P_1B_1$ and $P_2B_2$ respectively.

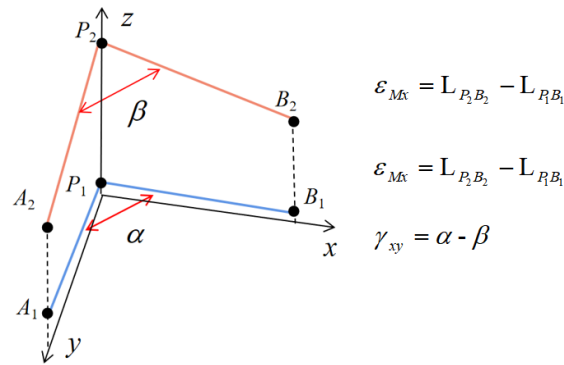

Fig.5 The strain calculation schematic diagram

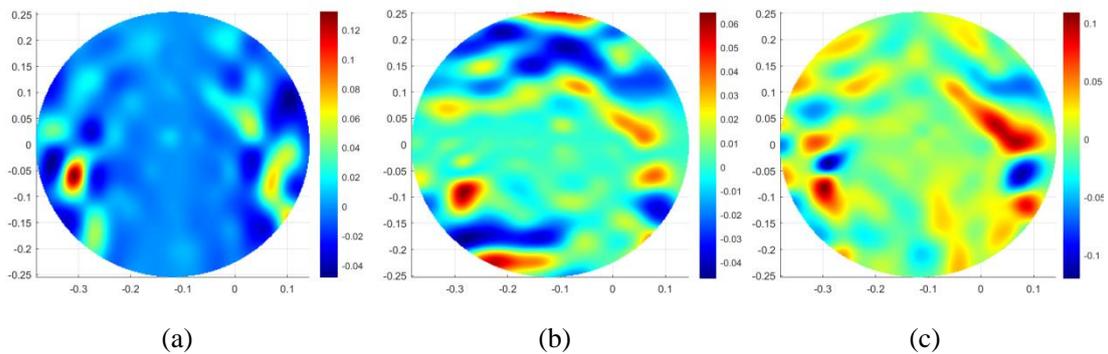

(a) (b) (c)

Fig.6 Strain in $x$ dimension and $y$ dimension

Matlab is used to calculate the membrane and shear strain. The results are shown in Fig.6. The membrane strain in $x$ and $y$ dimension of point P are 0.022 and -0.005 respectively, and the shear strain of point P is 0.017. As can be seen from equation (3), the bending stress scales with thickness. Assume the thickness is 0.5mm, 1 mm, 2 mm, 3 mm, 4 mm, the bending stress, the membrane stress and the shear stress are calculated by equation (6),(11),1(c), the results is shown in Fig.7. The Poisson's ratio corresponding to the minimum bending stress and membrane stress are -0.43, 0.12 respectively, and the shear stress decreases with Poisson's ratio increasing. The main principal stress calculated by equation (13) is shown in Fig.8. When the thickness is 0.5 mm, the Poission's ratio corresponding to the minimum main principal stress(short for "$v_{min}$") is 0.06, which is closer to $v_{Mmin}$, because the membrane stress is dominated. Along with the thickness increasing, the bending stress play more role in the main principle stress, and $v_{min}$ moves in negative direction and closer to $v_{B\min}$.

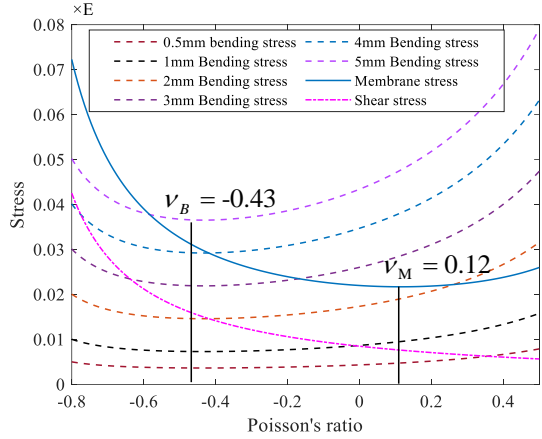 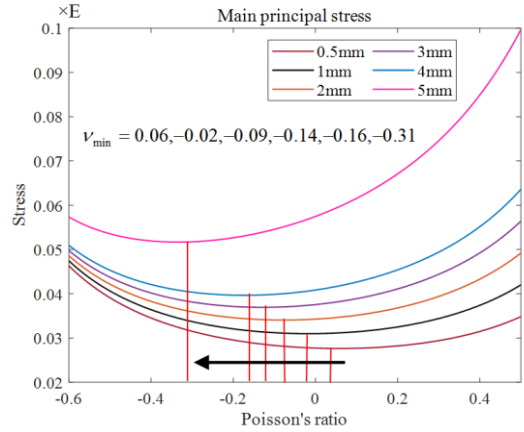

Fig.7 Stress vs. Poisson's ratio    Fig.8 Main principal stress vs. Poisson's ratio

## 5 Optimal Design of Reflector Surface

### 5.1 Lattice based structure

For the reflector surface with thickness exceed 1mm, Poisson's ratio $v$ is negative for minimum stress according to Fig.8. Traditional honey comb structure material is common used in reflector surface exhibits positive Poisson's ratio, a NPR lattice with elliptic voids introduced by Bertoldi and co-works[20] is proposed in this article, see Fig.9. The 2D structure consist of the matrix wherein the elliptical voids are arranged in periodic manner, and can be seen as an analogue of either the rotating square model with square-like elements of the matrix. The elliptical voids have two possible orientations, horizontal and vertical, and the elastic parameters in this two orientations are same[21]. The Poisson's ratio is designable by the adjustment of the geometry parameters $a$, $b$, $g$.

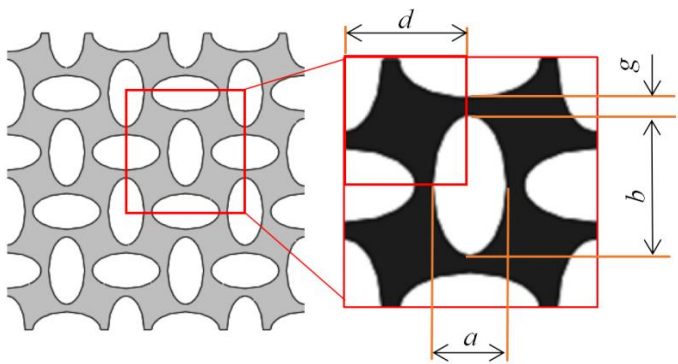 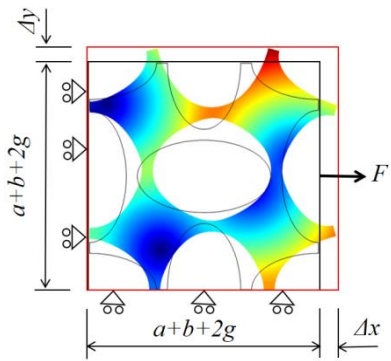

Fig.9 The elliptic voids structure NPR material    Fig.10 FE simulation model

During the optimization, the transmission loss should be taken into consideration, the diameter should be smaller than $\lambda/8$. As a example, for the X band, the wave length is about 30 mm, the parameter $a$ determined as 3.65mm. To lower the tensile stiffness, the parameter $g$ should be small, but that will increase the manufacturing difficulty and cost. To compromise this paradox, $g$ is supposed as 0.3mm. A FE model, as shown in Fig.10, is used to simulate the deformation and the effective Poisson's ratio $v_{eff}$ and the ratio of effective Young's modulus($E_{eff}$) to the Young's modulus of material($E_{mat}$) are calculated by the following way:

$$v_{eff} = -\frac{\varepsilon_y}{\varepsilon_x} = -\frac{\Delta y/(a+b+2g)}{\Delta x/(a+b+2g)} = -\frac{\Delta y}{\Delta x} \quad (16)$$

The material investigated is carbon fibre reinforced silicon(CFRS), for example, Young's modulus $E_{mat}$=500MPa, Poisson's ratio is 0.4. The simulation results is shown in Fig.11. With the ratio of $b/a$ increasing, the effective Poisson's ratio $v_{eff}$ increase.

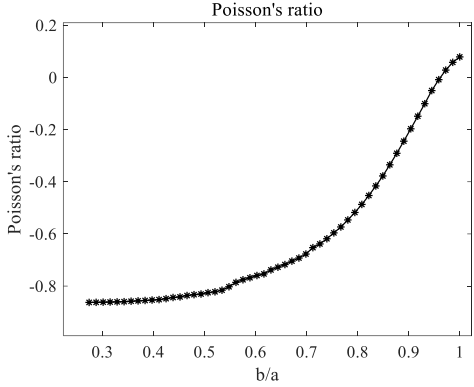
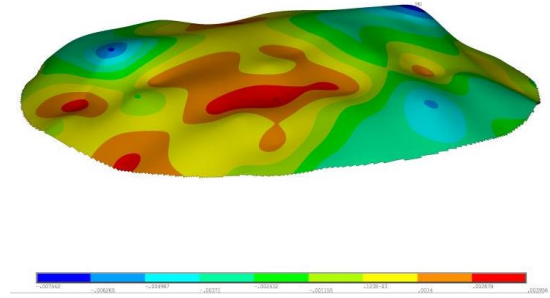

Fig.11 Effective Poisson's ratio vs. $b/a$     Fig.12 Simulated surface of reflector

## 5.2 Thickness of Reflector Surface

The thickness of reflector surface is a important parameters with an influence over out-of-plane deformation. As can be seen from equation(18), the bending stiffness is a cubic relationship of thickness. As mentioned above, the bending stiffness to tensile modulus should be large enough for achieving smooth reshaping. For this purpose, increasing thickness is a effective way which can significantly increase the bending stiffness, but it will lead the final displacement to lower local change in curvature and a more overall displacement.

$$D = \frac{Et^3}{12(1-v^2)} \quad (17)$$

To optimize the thickness, the discrete values of the variables investigated are given below:

$$v_{eff} = \{0.5mm, 1mm, 2mm, 3mm, 4mm, 5mm\}$$

According to stress analysis for vary thickness above in Fig.8, the Poisson's ratio corresponding to the minimum stress can be established. As the parameter $a$=3.65 mm and $g$=0.3mm of lattice are determined, the value of parameter $b$ can be determined according to the simulation results in Fig.11. So the parameters list as followed:

Table 1 the parameters of reflector surface

| $t$(mm) | 0.5 | 1 | 2 | 3 | 4 | 5 |
|---|---|---|---|---|---|---|
| $v_{eff}$ | 0.06 | -0.02 | -0.09 | -0.14 | -0.16 | -0.31 |
| $a$(mm) | | | 3.65 | | | |
| $g$(mm) | | | 0.3 | | | |
| $b$(mm) | 3.6 | 3.45 | 3.4 | 3.35 | 3.3 | 3.16 |

ANSYS is used to simulate the reconfigured reflector surface by the actuation of 225 actuators. And the actuators are located at points of the maximum Gaussian curvature in each even distribution region. One example of the simulated surface is shown in Fig.12. The surface accuracy of the deformed reflector is evaluated by the root mean square (RMS) error (18):

$$RMS = \sqrt{\frac{\sum_{i=1}^{n}(z_i^* - z_i)^2}{n}} \qquad (18)$$

RMS of reconfigured reflectors with vary thickness is calculated and shown in Fig.13, RMS decreases with thickness increasing firstly, and when thickness beyond a value( t=5mm in this case), the RMS decrease,see Fig.13(g). As the actuation force accompanying is also increasing significantly when thickness increase. Thus, at the constraint of the reconfiguration accuracy, small thickness is desirable.

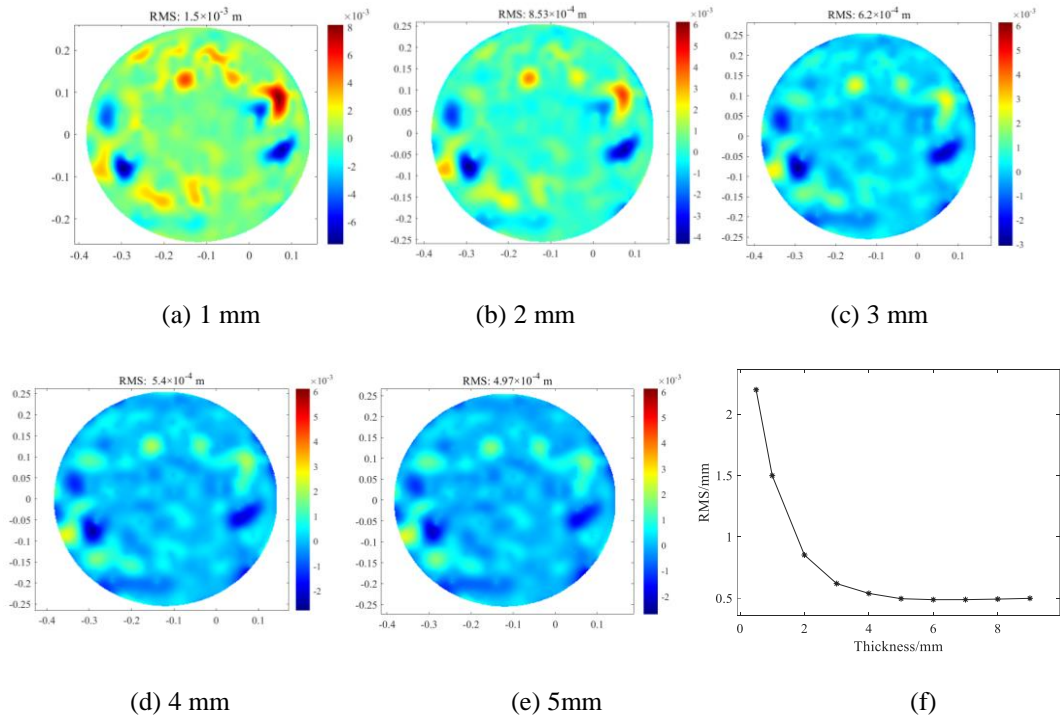

(a) 1 mm  (b) 2 mm  (c) 3 mm

(d) 4 mm  (e) 5mm  (f)

Fig.13 Error distribution between simulated reconfigured shape with thickness and original shape (a) 1mm (b) 2 mm (c) 3 mm (d) 4 mm (e) 5 mm (f) Curve of RMS vs.thickness

## 6 Conclusion

The investigation of Poisson's ratio for the material of mechanical reconfigurable reflector surface is given in this paper. The reshaping deformation is achieved though a combination of a change of curvature and in-plane strain. Based on the stress analysis, Poisson's ratio corresponding to the minimum stress are related to the thickness, the ratio of curvature change and the ratio of in-plane strains in two principal dimension. When curvature change in same direction (i.e. $\delta K_y/\delta K_x>0$), the Poisson's ratio corresponding to the minimum bending stress ($\nu_{Bmin}$) is negative, otherwise, $\nu_{Bmin}$ is positive, and when $\delta K_x \times \delta K_y=0$, $\nu_{Bmin}$ is zero. Similarly for the membrane stress, $\nu_{Mmin}$ are negative, positive and zero when $\delta\varepsilon_y/\delta\varepsilon_x>0$, $\delta\varepsilon_y/\delta\varepsilon_x<0$ and $\delta\varepsilon_y \times \delta\varepsilon_x=0$ respectively. And for the main principal stress which is found by superposition of bending and membrane stresses, when thickness is small, the membrane stress is dominated, $\nu_{min}$ is close to $\nu_{Mmin}$, and while thickness increasing, the bending stress is account for the main part, $\nu_{min}$ is close to $\nu_{Bmin}$.

A case of reconfigurable reflector surface is investigated in this paper. As for the demand of negative Poisson's ratio, a NPR lattice with elliptic voids is proposed, of which Poisson's ratio can be tailored by adjust the ellipticity. The determination of thickness should take accuracy and actuation force into consideration. Increasing the thickness can reduce the RMS, but also the actuation force increasing.